\documentclass{article}
\usepackage{spconf,amsmath,graphicx}
\usepackage{amsfonts}

\title{Decomposed Temporal Dynamic CNN: Efficient Time-Adaptive Network for Text-Independent Speaker Verification 
\\ Explained with Speaker Activation Map}
\name{Seong-Hu Kim, Hyeonuk Nam, Yong-Hwa Park\thanks{This research was financially supported by the Institute of Civil
Military Technology Cooperation funded by the Defense Acquisition Program Administration and Ministry of Trade, Industry and Energy of Korean government under grant No. UD10044TU.}}
\address{Department of mechanical Engineering, Korea Advanced Institute of Science and Technology, Korea}
\begin{document}
\ninept
\maketitle
\begin{abstract}
To extract accurate speaker information for text-independent speaker verification, temporal dynamic CNNs (TDY-CNNs) adapting kernels to each time bin was proposed. However, model size of TDY-CNN is too large and the adaptive kernel's degree of freedom is limited. To address these limitations, we propose decomposed temporal dynamic CNNs (DTDY-CNNs) which forms time-adaptive kernel by combining static kernel with dynamic residual based on matrix decomposition. Proposed DTDY-ResNet-34(×0.50) using attentive statistical pooling without data augmentation shows EER of 0.96\%, which is better than other state-of-the-art methods. DTDY-CNNs are successful upgrade of TDY-CNNs, reducing the model size by 64\% and enhancing the performance. We showed that DTDY-CNNs extract more accurate frame-level speaker embeddings as well compared to TDY-CNNs. Detailed behaviors of DTDY-ResNet-34(×0.50) on extraction of speaker information were analyzed using speaker activation map (SAM) produced by modified gradient-weighted class activation mapping (Grad-CAM) for speaker verification. DTDY-ResNet-34(×0.50) effectively extracts speaker information from not only formant frequencies but also high frequency information of unvoiced phonemes, thus explaining its outstanding performance on text-independent speaker verification.
\end{abstract}

\section{Introduction}
Text-independent speaker verification uses utterances with random texts composed of different phoneme sequence and acoustic characteristics, but the same deep learning models and methodologies as text-dependent tasks are utilized. In fact, speaker embeddings and verification performance depend on phonemes \cite{RN1, RN2, RN3}, so we inferred that conventional text-independent speaker verification methods do not consider the effect of random texts with different phoneme sequence. Thus, we proposed temporal dynamic models in previous works to extract accurate frame-level speaker information depending on phonemes by adapting convolutional neural network (CNN) kernels with the time bins of input utterances \cite{RN4, RN5}, and there are other recent works with similar approaches \cite{RN6, RN7}. 

Among them, temporal dynamic convolutional neural network (TDY-CNN) \cite{RN4} applies adaptive kernels defined by weighted summation of basis kernels and extracts accurate frame-level speaker information. Its result proved the advantage of temporal content-adaptive models that consider phoneme variations in random texts on text-independent speaker verification. However, generating adaptive kernels using \(K\) basis kernels may cause a constraint as adaptive kernels can only have up to \(K\) degrees of freedom, thus the diversity of adaptive kernels might not be sufficient to consider various phonemes. Simple solution to this is to use more basis kernels, but it will increase the number of model parameter exponentially. In addition, this approach have several limitations such as lack of compactness and challenge in joint optimization of attention weights and basis kernels \cite{RN4,RN8}. Similar limitations on large model size and  model optimization have also been raised in large-scale vanilla convolution, which are dealt by applying matrix decomposition to convolution kernel \cite{RNdecomp1,RNdecomp2}. Similar approach was applied to dynamic convolution by applying matrix decomposition to dynamic convolution kernel \cite{RN9}.
Thus, by applying matrix decomposition to TDY-CNN, we propose decomposed temporal dynamic convolutional neural network (DTDY-CNN) to address the limitations of TDY-CNN and improve text-independent speaker verification performance. 

In addition, to explain how TDY-CNN extracted speaker embeddings depending on phonemes, an analysis on how kernels adapt to different phonemes was introduced in \cite{RN4}. The result has shown that phoneme groups sharing similar acoustic characteristics use similar kernels of TDY-CNN. The analysis has deepened our understanding in TDY-CNN, but it is hard to explain how the temporal dynamic model utilized phoneme-dependent kernels to extract speaker information from utterances with random phoneme sequence. Such problems have been commonly raised on artificial intelligence(AI), so explainable AI \cite{RN10} has been proposed to explain and interpret black-box models including deep learning models. In particular, gradient-weighted class activation mapping (Grad-CAM) \cite{RN11} is proposed to provide visual explanation on various CNN-based models. We apply Grad-CAM on text-independent speaker verification models and explain whether temporal dynamic models adapt to various phonemes.

In this paper, we propose decomposed temporal dynamic convolutional neural networks (DTDY-CNNs) for text-independent speaker verification. Main contributions of this work are as follows: 
\begin{enumerate}
\item Matrix decomposition is applied to temporal dynamic kernels to ensure that adaptive kernels fully consider time-varying phonemes with more efficient manner.
\item We explained how DTDY-CNN extracts speaker embeddings depending on phonemes by analysis of frame-level embeddings and speaker activation maps (SAMs). 
\end{enumerate}
The remainder of the paper is organized as follows. Section 2 introduces DTDY-CNN for speaker verification. Section 3 describes baseline models and training details with evaluation metrics. Section 4 shows the experiment results and analysis. Finally, Section 5 presents conclusions.

\section{Decomposed Temporal Dynamic Convolutional Neural Network}

For effective extraction of speaker information from various utterances, we propose decomposed temporal dynamic convolution that applies matrix decomposed adaptive convolution depending on time bins as follows: 
\begin{align}
    {y}(f,t) &=W(t)*x(f,t) \label{eq1},\\
    W(t) &={W}_{0}+P\Phi (t) Q^{T},\label{eq2}
\end{align}
where \(*\) denotes convolution. A temporal adaptive kernel \(W(t)\in\mathbb{R}^{{C}_{out}\times{C}_{in}\times{k}\times{k}}\) is convoluted with the input \(x(f,t)\in\mathbb{R}^{{C}_{in}\times{F}\times{T}}\) and the result is output \(y(f,t)\in\mathbb{R}^{{C}_{out}\times{F'}\times{T'}}\), where \(C_{in}\) and \(C_{out}\) are the numbers of input and output channels, \(k\) is kernel size, \(F\) and \(T\) are input feature shapes, \(F'\) and \(T'\) are output feature shapes, respectively. \(W(t)\) is composed of static kernel \(W_0\) and dynamic residual \(P\Phi Q^T\). The input is compressed by \(Q\in\mathbb{R}^{{C}_{in}\times L\times k\times k}\) into a lower dimension \(L\) so that \(Q^{T}*x\in\mathbb{R}^{L\times{F'}\times{T'}}\). Temporal dynamic matrix \(\Phi (t) \in \mathbb{R}^{L \times L}\) is a linear transformation matrix in the \(L\)-dimensional latent space, and there are different linear transformations for each time bin \(t\). The temporal dynamic matrix is obtained by applying two fully-connected layers with squeeze-and-excitation \cite{RNSE} to a feature generated by concatenating average pooling features along channel \(C_{in}\) and frequency \(F\) dimensions. The concatenate feature of each time bin \(t\) are squeezed to \(({C}_{in}+F)/r\) dimension with reduction ratio \(r\) and expanded to \(L^2\) dimension. We set latent space dimension \(L\) as  \(\sqrt{2C_{in}+2C_{out}}\). The dynamically fused result is expanded by \(P \in \mathbb{R}^{C_{out} \times L}\) into the output channel dimension \(C_{out}\). The decomposed temporal dynamic convolution applies temporal dynamic transformation adaptive to input in the latent space containing major information about speakers.

\begin{figure}[t]
  \centering
  \includegraphics[width=\linewidth]{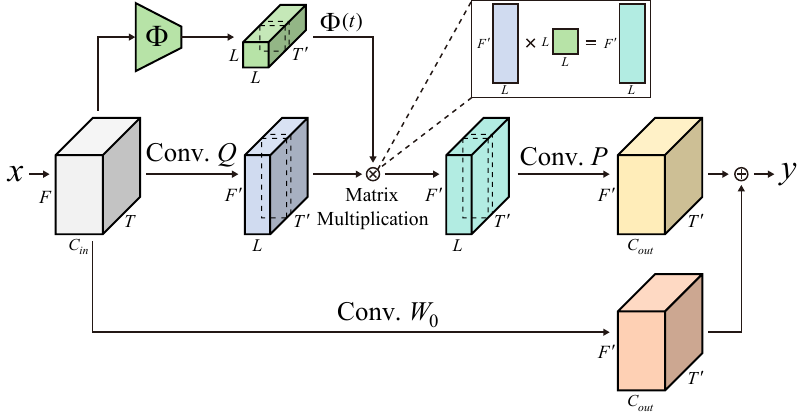}
  \caption{Structure of decomposed temporal dynamic convolution. It aggregates convolution results of static kernel \(W_0\) and dynamic residual \(P \Phi(t) Q^T\).}
  \label{fig1}
  \vspace{-0.1cm}
\end{figure}

In decomposed temporal dynamic convolution, we implement the dynamic residual of the temporal adaptive kernel using two convolution layers with a matrix multiplication in between. The convolution output with this residual kernel is added to convolution output of static kernel as shown in Figure \ref{fig1}. \(Q\) and \(P\) are equivalent to \(k\times k\) and \(1\times 1\) 2D convolution, respectively. The temporal dynamic matrix is applied to each time bin using matrix multiplication between \(Q\) and \(P\) layers. The decomposed temporal dynamic convolution is applied by replacing the vanilla convolution in speaker validation models, and this networks are called decomposed temporal dynamic convolutional neural networks (DTDY-CNNs). We expect DTDY-CNN to consider phonemes by securing the diversity of adaptive kernels using temporal dynamic matrix in \(L\)-dimensional latent space with fewer parameters.

\section{Experiment Setup and Details}
\subsection{Input representations}
2-second-long segments are randomly extracted from training utterances. The segments are converted into 64-dimensional log Mel-spectrograms using 25ms hamming window with 10ms step and the number of fast Fourier transform 512. Mean and variance normalization is performed on every frequency bin. The Mel-spectrograms with a size of 64×200 are used for training.

\subsection{Baseline architecture and loss}
As residual networks have been shown to be effective on speaker recognition, we use ResNet-34 as a baseline model \cite{RN13}. The channels of ResNet-34 are reduced to a quarter (ResNet-34(×0.25)) and half (ResNet-34(×0.50)) for computational cost reduction. Frame-level features are aggregated using average pooling to apply uniform attention and backward gradients to each frame-level feature. Decomposed temporal dynamic convolution replaces all vanilla convolution of ResNet except for the first convolution layer that extracts global features \cite{RN8}. We called ResNet using decomposed temporal dynamic convolution as DTDY-ResNet.

Models are trained using a loss function combining the angular prototypical loss with the vanilla softmax loss. This method demonstrates better verification performance than using each or other combinations of the loss functions \cite{RN14}.

\subsection{Implementation details and datasets}
The models are trained for 200 epochs using the Adam optimizer with weight decay \(5\times10^{-5}\). An initial learning rate is \(10^{-3}\) and it decreases by a factor of 0.75 every 10 epochs. Batch normalization is used with a fixed batch size of 1024 consisting of 512 mini-batch from 2 utterances for angular prototypical loss. No data augmentation is performed during training. We use 5994 speakers of Voxceleb2 development set for training \cite{RN15}, and the models are validated on Voxceleb1 test set \cite{RN16}. 

\subsection{Evaluation metrics}
Ten 4-second-long segments are sampled with the same intervals from each test segment, and 100 similarities between every pair of segments are computed. Mean of the 100 similarities is used as the final pairwise score \cite{RN14, RN15}. Based on the average similarity score between speakers, we calculate equal error rate (EER) and the minimum value of the cost function \(C_{det}\) with parameters \(C_{miss}=1\), \(C_{fa}=1\), and \(P_{target}=0.05\) \cite{RN16, RN17} as evaluation metrics of verification.

\section{Result and Analysis}
\subsection{Text-independent speaker verification using DTDY-CNN}
\subsubsection{Reduction ratio of temporal dynamic matrix generator}
Reduction ratio \(r\) indicates how much speaker and phoneme information are squeezed between layers, so \(r\) is directly related to the model complexity and speaker verification performance. We compared the text-independent speaker verification performance of DTDY-ResNet-34(×0.25) with different reduction ratios to optimize decomposed temporal dynamic convolution. The models are tested for three cases (\(r\) = 1/4, 1/8, 1/16) as shown in Table \ref{tab1}. DTDY-ResNet-34(×0.25) with 1/8 of \(r\) yield the best performance, and the error rate was degraded as 1/4 and 1/16 of \(r\). High reduction ratio can occur unstable optimization and overfitting while the representation power of the model increases. Low reduction ratio reduces computational costs, but it compromises the performance by squashing important information (speakers, phonemes, etc.). Thus, we chose decomposed temporal dynamic convolution with \(r\) = 1/8 as the optimal structure for text-independent speaker verification.

\begin{table}[t]
\vspace{-0.2cm}
    \caption{Speaker verification performance of decomposed temporal dynamic models with different reduction ratio.}
    \vspace{0.1cm}
    \label{tab1}
    \centering
\begin{tabular}{c|c|c|c}
\hline
\bf{DTDY-ResNet-34(×0.25)}   & \bf{\#Parm} & \bf{EER(\%)}  & \bf{MinDCF} \\ \hline
\(r\) = 1/4  & 4.42M  & 1.70 & \bf{0.129}  \\
\(r\) = 1/8  & 3.29M   & \bf{1.59} & 0.130  \\ 
\(r\) = 1/16 & 2.73M  & 1.88 & 0.139  \\ \hline
\end{tabular}
\end{table}

\begin{table}[t]
\vspace{-0.3cm}
    \caption{Speaker verification performance of the networks without data augmentation.}
    \vspace{0.1cm}
    \label{tab2}
    \centering
\begin{tabular}{c|c|c|c}
\hline
\bf{Network}   & \bf{\#Parm} & \bf{EER(\%)}  & \bf{MinDCF} \\ \hline
\multicolumn{1}{l|}{ResNet-34(×0.25)} & 1.86M  & 2.05 & 0.153  \\ 
\multicolumn{1}{l|}{TDY-ResNet-34(×0.25)}  & 8.66M   & 1.70 & \bf{0.122}  \\ 
\multicolumn{1}{l|}{DTDY-ResNet-34(×0.25)}  & 3.29M  & \bf{1.59} & 0.130  \\ \hline
\multicolumn{1}{l|}{ResNet-34(×0.50)} & 6.37M  & 1.66 & 0.131  \\ 
\multicolumn{1}{l|}{TDY-ResNet-34(×0.50)}  & 33.5M   & 1.46 & 0.116  \\ 
\multicolumn{1}{l|}{DTDY-ResNet-34(×0.50)}  & 12.0M  & \bf{1.37} & \bf{0.103}  \\
\multicolumn{1}{l|}{\quad+ASP}  & 13.6M  & \bf{0.96} & \bf{0.086}  \\ \hline \hline
\multicolumn{1}{l|}{ResNet-50 \cite{RN15}} & 67.0M  & 3.95 & 0.429  \\ 
\multicolumn{1}{l|}{Thin ResNet-34 \cite{RN17}}  & 12.4M   & 2.87 & 0.310  \\ 
\multicolumn{1}{l|}{ResNet-Q/SAP \cite{RN18}} & 1.40M  & 1.47 & 0.119  \\
\multicolumn{1}{l|}{ResNet-H/ASP \cite{RN18}} & 8.00M  & 1.21 & 0.098  \\
\multicolumn{1}{l|}{ECAPA-TDNN \cite{RN19}}  & 15.6M  & 1.20 & 0.101  \\ 
\hline
\end{tabular}
\vspace{-0.2cm}
\end{table}

\subsubsection{Text-independent speaker verification results}
The text-independent speaker verification performances of DTDY-ResNets are compared to baseline models and TDY-ResNets using six basis kernels \cite{RN4} as shown in Table \ref{tab2}. DTDY-ResNet-34 models outperform ResNet-34 models which use vanilla convolution and TDY-ResNet-34 models using temporal dynamic convolution. Especially, DTDY-ResNet-34(×0.50) not only shows  better performance as EER of 1.37\% than TDY-ResNet-34(×0.50) with EER of 1.46\%, but also reduced the number of model parameters by 64\%. The previous networks utilized average temporal pooling to see only the affection of the proposed decomposed temporal dynamic convolution, and it is necessary to consider another temporal pooling method.  We applied attentive statistical pooling (ASP) \cite{RNASP} to DTDY-ResNet-34(×0.50), and DTDY-ResNet-34(×0.50) using ASP shows the best performance as EER of 0.96\%. Thus, DTDY-CNNs  improve speaker verification performance with fewer model parameters than TDY-CNNs due to its compact and efficient dynamic structure based on matrix decomposition.

In addition, DTDY-ResNets are compared to state-of-the-art networks \cite{RN15, RN17, RN18, RN19} trained with Voxceleb2 without data augmentation and tested with Voxceleb1, same as our experiment condition. DTDY-ResNet-34(×0.50) using ASP outperforms even the state-of-the-art networks with EER of 0.96\%. From the speaker verification results, we conclude that DTDY-CNNs are more optimal for text-independent speaker verification than TDY-CNNs and the state-of-the-art networks.

\subsection{Comparison of TDY-CNN and DTDY-CNN}
\subsubsection{Frame-level speaker embeddings  on different phonemes}
\begin{figure}[t]
  \centering
  \includegraphics[width=0.85\linewidth]{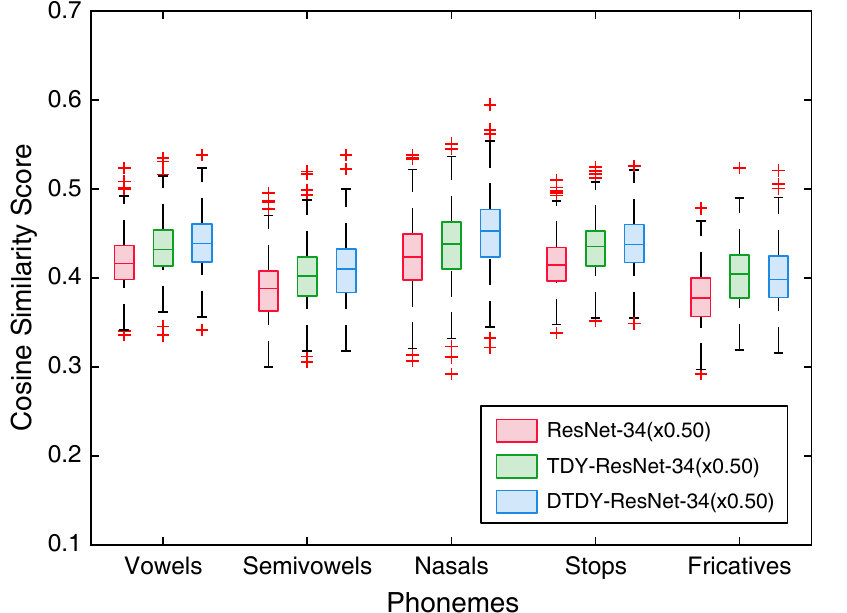}
  \vspace{-0.2cm}
  \caption{Distribution of cosine similarity score between utterance-level and frame-level speaker embeddings depending on phonemes.}
  \label{fig2}
  \vspace{-0.3cm}
\end{figure}

Our hypothesis was that DTDY-CNNs compute accurate frame-level speaker embeddings depending on phonemes from utterances with random texts compared to TDY-CNNs. To verify this, we compared cosine similarity between utterance-level and frame-level speaker embeddings depending on the phoneme groups within a same speaker using TIMIT dataset \cite{RN20}. The utterance-level speaker embedding is derived by averaging 9 utterance-level embeddings, excluding the target utterance from which frame-level speaker embeddings are extracted \cite{RN3, RN5}. The frame-level speaker embeddings are extracted by removing the average pooling layer of models. The  cosine similarity score distribution of TDY-ResNet-34(×0.50) and DTDY-ResNet-34(×0.50) were calculated and compared depending on five phoneme groups as shown in Figure \ref{fig2}. 

For all phoneme groups, TDY-ResNet-34(×0.50) has higher cosine similarity scores between utterance-level and frame-level embeddings than ResNet-34(×0.50) using vanilla convolution, identical to the previous study result \cite{RN4}. The cosine similarity scores of DTDY-ResNet-34(×0.50) outperform that of TDY-ResNet-34(×0.50) for all  phoneme groups except for fricatives, and similar scores are obtained for fricatives. This result indicates that more speaker information is contained in the frame-level embeddings extracted by DTDY-ResNet-34(×0.50) than in the case of ResNet-34(×0.50) and TDY-ResNet-34(×0.50). Therefore, the proposed DTDY-CNN extracts the relatively accurate frame-level speaker embeddings for all phoneme groups phonemes compared to TDY-CNN.

\begin{figure*}[t!]
    \centering 
    \includegraphics[width=0.48\linewidth]{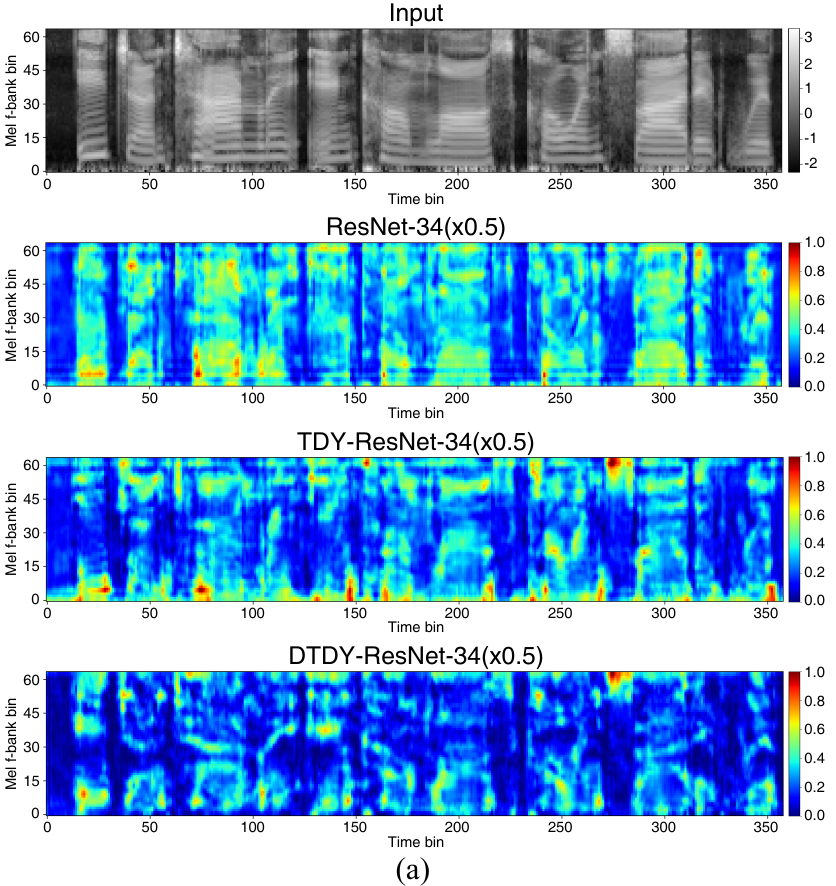} \quad
    \includegraphics[width=0.48\linewidth]{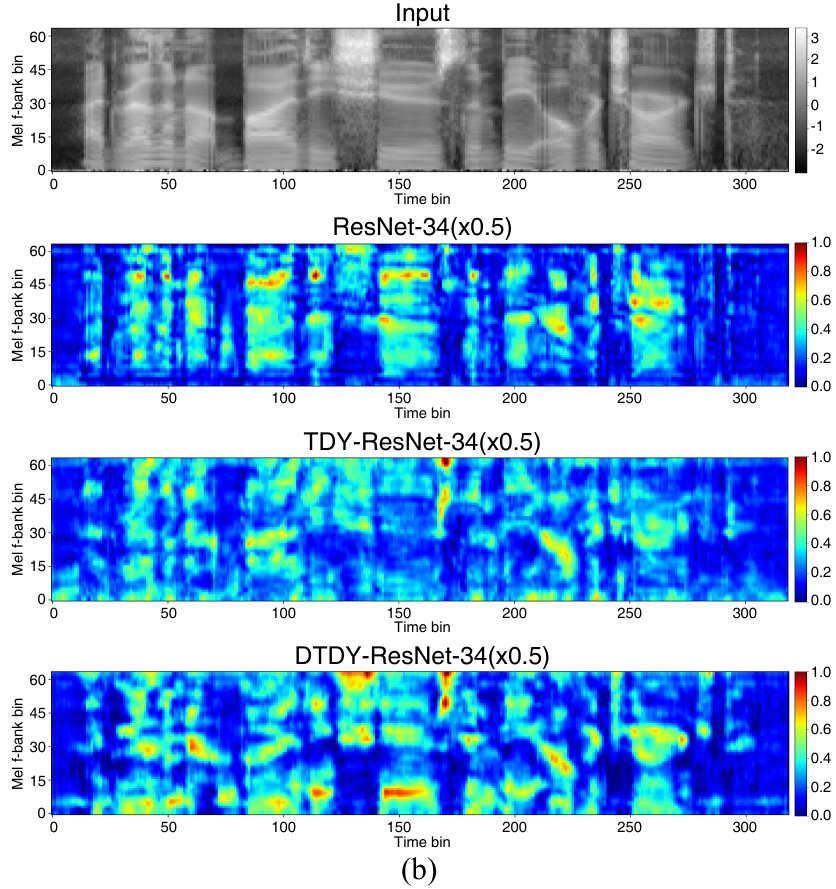}
     \vspace{-0.3cm}
    \caption{Speaker activation maps of first 2 residual block in ResNet-34(×0.5), TDY-ResNet-34(×0.5), and DTDY-ResNet-34(×0.5) for (a) SX364 by FEDW0 and (b) SX298 by MTAS1.}
    \label{fig3}
    \vspace{-0.3cm}
\end{figure*}

\subsubsection{Explanation of temporal dynamic models using Grad-CAM}
DTDY-CNN extracts accurate speaker information by adaptive convolution kernels depending on phonemes using less model parameters. Moreover, to secure the validity of DTDY-CNN for text-independent speaker verification, it is necessary to objectively evaluate the ability of DTDY-CNN to activate key information of utterance compared to TDY-CNN. In this section, we use Grad-CAM \cite{RN11} to analyze from which part of utterances the dynamic models extract speaker information. As Grad-CAM produces a coarse localization map for any target using target-specific gradients flowing into the final convolutional layer, it is widely used to explain the behavior of trained deep neural networks. 

However, there are two problems in applying Grad-CAM to speaker verification models: (i) unclear target for speaker verification network architecture and (ii) inappropriateness of target localization map on speaker verification. For (i), it is difficult to clearly define the target of speaker verification models from which we calculate the gradients because the purpose of model is speaker embedding extraction. To solve this problem, we applied Grad-CAM to speaker recognition models trained by attaching a classifier after the pre-trained speaker verification models that extract speaker embeddings. The classification layer was trained to recognize 630 speakers from speaker embeddings, using TIMIT dataset \cite{RN20}. 10 utterances for each speaker are divided into 2 utterances for test and 8 utterances for training. For (ii), as any part of the utterance belongs to one specific speaker, there is no point localizing the target map. Rather, it is necessary to visualize how the speaker verification model extracts speaker information from different frequency patterns of various phonemes at each time point. As the information regarding phonemes would be taken in earlier layer and the later layers would be most likely to contain only speaker-related information, we produced the speaker activation map (SAM) using gradients flowing into the first convolution layers. Therefore, we modified Grad-CAM to visualize SAM for text-independent speaker verification models as mentioned above. 

SAMs of ResNet-34(×0.50), TDY-ResNet-34(×0.50), and DTDY-ResNet-34(×0.50) for utterance SX364 by speaker FEDW0 and utterance SX298 by speaker MTAS1 are compared as shown in Figure \ref{fig3}. ResNet-34(×0.50) activates wide frequency region of uttered section and vague outline of each phoneme. On the contrary, TDY-ResNet-34(×0.50) and DTDY-ResNet-34(×0.50) activate the low-frequency section for voiced sounds and the high-frequency section for fricative-like sounds. These models capture formant frequency regions of phonemes more clearly, so the kernels of TDY-ResNet-34(×0.50) and DTDY-ResNet-34(×0.50) extracted speaker information depending on phonemes. However, there are differences between two temporal dynamic models. TDY-ResNet-34(×0.50) activates vague frequency bands, whereas DTDY-ResNet-34(×0.50) focus more on formant frequencies. Considering that speaker is recognized by one’s voice tone with formant frequencies, DTDY-CNN shows more speaker specific-characteristics to consider detailed features of phonemes than TDY-CNN does.

\section{Conclusion}
In this paper, we proposed decomposed temporal dynamic convolutional neural network (DTDY-CNN) to extract accurate speaker information from various phonemes by adapting kernels to each time bin for text-independent speaker verification. The adaptive kernel of decomposed temporal dynamic convolution is composed of static kernel and dynamic residual based on matrix decomposition for more efficient adaptive kernels. Proposed DTDY-ResNet-34(×0.50) using ASP showed best verification performance as EER of 0.96\%. DTDY-ResNet-34(×0.50) reduced the model parameters by 64\% and extracted more accurate frame-level speaker embeddings compared to TDY-ResNet-34(×0.50). In addition, DTDY-ResNet-34(×0.50) was analyzed using speaker activation map (SAM) produced by modified Grad-CAM to explain how DTDY-CNN extracts speaker information from utterances. TDY-ResNet(×0.50) activated vague formant frequency band regions, whereas DTDY-ResNet(×0.50) extracted speaker information from more specific regions of formant frequencies depending on phonemes. Therefore, DTDY-CNN considers detailed features of phonemes maintaining speaker verification performance even with a small number of parameters. Both results and analysis indicate that DTDY-CNN improves text-independent speaker verification performance by effectively extracting speaker information from various phonemes compared to TDY-CNN.

\vfill
\pagebreak

\bibliographystyle{IEEEbib}
\bibliography{paper.bib}

\end{document}